# Current-Confined Formation of a Buried Ge/Si Free Surface for Defect Filtering

Youcef A. Bioud [1*], Meriem Bouchilaoun [2], Waldemar Schreiber [1], Redouane Amrar [2,3], Gilles Patriarche [4], Ali Soltani [2,3], Stefan Janz [1*]

*[1] Division Photovoltaics, Fraunhofer Institute for Solar Energy Systems (ISE), Heidenhofstraße 2, 79110 Freiburg, Germany.*
*[2] Institut Interdisciplinaire d'Innovation Technologique (3IT), Université de Sherbrooke, 3000 Boulevard de l'université, Sherbrooke, J1K OA5 Québec, Canada*
*[3] Laboratoire Nanotechnologies Nanosystèmes (LN2) - IRL3463, CNRS, Université de Sherbrooke, INSA Lyon, École Centrale de Lyon, Université Grenoble Alpes, Sherbrooke, J1K 0A5 Québec, Canada.*
*[4] Université Paris-Saclay, CNRS, Centre de Nanosciences et de Nanotechnologies, 91120 Palaiseau, Paris, France.*

**Email: y.bioud@usherbrooke.ca; stefan.janz@ise.fraunhofer.de*

**Abstract.**

Lattice-mismatched heteroepitaxy is limited by misfit dislocations, threading defects, and residual strain generated at buried interfaces. Here, we show that current-confined electrochemistry transforms the Ge/Si heterointerface into a laterally continuous nanometric free surface while preserving the Ge epilayer. A low-doped Ge layer on a conductive p-type Si substrate redirects anodization current toward the misfit-dislocation-rich interface. Numerical modelling including charge-transfer kinetics shows that this confinement arises from hole-limited charge transfer and, together with electrochemical calibration, defines a process window that combines interfacial current confinement with controlled Si porosification. STEM reveals an approximately 15 Å separation, while EDX shows localized interfacial oxygen enrichment. After soft annealing, geometric phase analysis and plan-view TEM show a modified strain-relaxation pathway and strongly suppressed threading-dislocation emergence. Raman spectroscopy indicates a reduction in residual tensile strain from approximately 0.10% to 0.05%, whereas spreading resistance profiling confirms preservation of the Ge carrier profile. Current-confined electrochemistry therefore provides a controllable post-epitaxial route for buried-interface decoupling, defect filtering, and strain control in lattice-mismatched semiconductor heterostructures.

**Keywords.**

Ge-on-Si, heteroepitaxy, electrochemical etching, interface engineering, dislocation reduction, strain relaxation, two-dimensional free surface.

Integrating lattice-mismatched semiconductor heterostructures remains central to optoelectronics, photonics, sensing, photovoltaics, and microelectronics.[1−3] Ge-on-Si is a prominent example. In nanoelectronics, it enables reconfigurable Ge field-effect transistors[4] and $Si_{1-x}Ge_x$ nanosheet devices[5]; in photonics, it underpins Ge-on-Si photodetectors[6]. Ultra-thin-body Si/Ge transistors further require Ge growth on SOI at a low thermal budget, which ultralow-temperature epitaxy has recently made possible[7]. In Ge/Si, the approximately 4.2% lattice mismatch and the difference in thermal-expansion coefficients generate an interfacial misfit-dislocation network once the critical thickness is exceeded, together with threading dislocations (TDs) that propagate into the active layer.[8-10] These defects degrade carrier transport and lifetime, optical response, and leakage behavior.[11-13] Conventional approaches—including compositional grading, cyclic annealing, epitaxial lateral overgrowth, selective-area growth, compliant substrates, and aspect-ratio trapping—can lower TD density. However, they often require complex or high-temperature processing, introduce wafer bowing or discontinuous active areas, or leave residual defect propagation.[12,14-19] Post-epitaxial electrochemical transformation offers a complementary route because internal free surfaces can redirect defect motion after growth. Nanovoids introduced within Ge/Si have been shown to attract dislocations and promote annihilation of their threading segments during annealing, enabling virtual-substrate formation.[20–26] Our previous work extended this concept to a sub-nanometric two-dimensional free surface (2DFS) at the Ge/Si interface and demonstrated its ability to decouple the epilayer and suppress defect propagation.[27] The unresolved question is how such a buried free surface nucleates selectively, propagates along the heterointerface, and remains stable without consuming the Ge overlayer. Here, we show that electrochemical current confinement governs this process and defines the material and operating conditions required for selective 2DFS formation. Going further than establishing the 2DFS itself, we identify hole-limited charge transfer as the origin of current confinement, relate interfacial selectivity to hole accumulation at the heterojunction, and derive quantitative design rules for doping, current density, and mesa size. We further link this controlled interfacial transformation to strain redistribution and defect filtering, using only a soft post-process anneal compatible with low-thermal-budget Ge integration.

Selective 2DFS formation requires dissolution to initiate at the buried Ge/Si interface rather than at the top Ge surface or uniformly within Si. Ge/Si mesa arrays were fabricated from a 3-µm-thick p-type Ge epilayer on p-type Si; fabrication and electrochemical procedures are detailed *in Sec. S1 of the supplementary material*. The Ge surface was protected by an HF-resistant resist, and bipolar anodic/cathodic pulses, previously shown to stabilize semiconductor porosification,[28,29] were applied in a 5:1 HF (49%):anhydrous-ethanol electrolyte. The electrical design combined low-doped Ge ($10^{15}$ cm$^{-3}$) with highly doped Si ($10^{18}$ cm$^{-3}$). Because anodic dissolution of p-type semiconductors consumes holes supplied at the surface, this contrast favors charge transfer at Si surfaces. The mesa architecture exposes the sidewall and the buried junction to the electrolyte while the protected top surface suppresses a competing vertical reaction path. This geometry allows anodization current entering through Si to reach the heterointerface and propagate laterally beneath Ge, which is essential for converting an extended buried region rather than producing isolated surface pits.

The mesa geometry and the axisymmetric numerical model are shown in Figs. 1(a) and 1(b); the model, its parameters, and a sensitivity analysis are given in Sec. S2 of the supplementary material. Bulk conductivities follow from the nominal doping, and linearized charge-transfer kinetics, with a resistance scaling inversely with the surface hole concentration, apply at all semiconductor/electrolyte boundaries.[30,31] A purely ohmic boundary would direct 77% of the anodic current through the Ge sidewall [Fig. 1(d)], which is inconsistent with the intact Ge layer observed below. With charge-transfer kinetics, this fraction falls to 0.07% [Fig. 1(e)]: the current bypasses the low-doped Ge and reaches Si and the Ge/Si boundary, over the investigated electrolyte-conductivity range, without requiring an assumed interfacial conducting layer (Sec. S2.2). Confinement therefore originates from hole-limited charge transfer set by the doping contrast, rather than from bulk conduction. This confinement establishes the electrical basis for selective buried-interface transformation. Planar separation further requires the heterointerface to be more reactive than the surrounding Si. A one-dimensional Poisson calculation [Fig. 1(c); Sec. S2.4] shows that the Ge/Si valence-band offset accumulates holes on the Ge side up to ~$10^{19}$ cm$^{-3}$, about ten times the substrate concentration, within a few nanometres, while the adjacent Si is depleted. Impurity trapping at dislocation cores can locally increase the effective carrier concentration and shift the Fermi level, thereby lowering the dissolution barrier,[32] while the associated lattice distortion and stress field can increase chemical reactivity.[33] Such an enhanced interfacial reactivity concentrates the current at the junction without increasing Ge attack, so that the front advances along the interface faster than the underlying Si is porosified. Once the interface opens at the sidewall, the model predicts that the current feeding the front flows through the thin interfacial gap and decays over a characteristic length $\lambda = (\sigma_{el} h r_{ct})^{1/2}$ of a few micrometres [Fig. 1(f); Sec. S2.3]; complete decoupling is therefore expected for mesas of comparable width, consistent with the 3 × 3 $\mu m^2$ geometry. In this picture, the macroscopic doping contrast determines where charge is delivered, whereas the interfacial band and defect structure defines the sites of preferential dissolution. Together, these effects confine etching to the defective interface, enabling the removal of a nanometric interfacial layer while preserving the micrometer-thick Ge overlayer.

Current confinement alone is insufficient because excessive Si dissolution or porosity destabilizes the Ge layer. We therefore calibrated Si etch rate and porosity as functions of current density and substrate carrier concentration [Figs. 1(g) and 1(h)]. Both quantities increase with current density and with Si doping, reflecting enhanced conductivity, hole availability, and anodic charge transfer. Porosity was extracted from infrared Fabry–Pérot fringes using an effective-medium model (*Sec. S3 of the supplementary material*).[34] Above approximately 65% porosity, the Ge overlayer progressively peels from the substrate. The $10^{18}$ $cm^{-3}$ Si substrate provides the best compromise between efficient interfacial transport and moderate porosification: $10^{19}$ $cm^{-3}$ approaches the unstable regime, whereas $10^{17}$ $cm^{-3}$ produces a weaker electrochemical response. This model-supported operating window agrees with the simulated selectivity map [Fig. 1(i)], in which the fraction of current passing through Ge remains below 1% for $N_{Si}/N_{Ge} \gtrsim 60$. Its experimental boundaries require comparative Ge/Si interface imaging across processing conditions. Accordingly, 3 × 3 $\mu m^2$ mesas were treated at ±50 mA $cm^{-2}$ using 1-s anodic and cathodic pulses for a total of 1 min, yielding less than 40% porosity. Under these conditions, low-doped Ge remains comparatively passive, conductive Si sustains most of the current, and dissolution is concentrated beneath Ge and along the defective heterointerface. The behavior is consistent with established divalent and tetravalent Ge dissolution and sequential Si fluorination pathways in HF media, summarized in *Sec. S4 of the supplementary material.* [29,35] Figure 1 therefore defines a combined electrical-electrochemical operating window: the doping contrast localizes current, while the pulse conditions restrict porosification below the mechanical-instability threshold. The selected conditions deliberately balance lateral interfacial transformation against vertical Si consumption, because rapid pore growth eventually removes the mechanical support required to retain the Ge mesa.

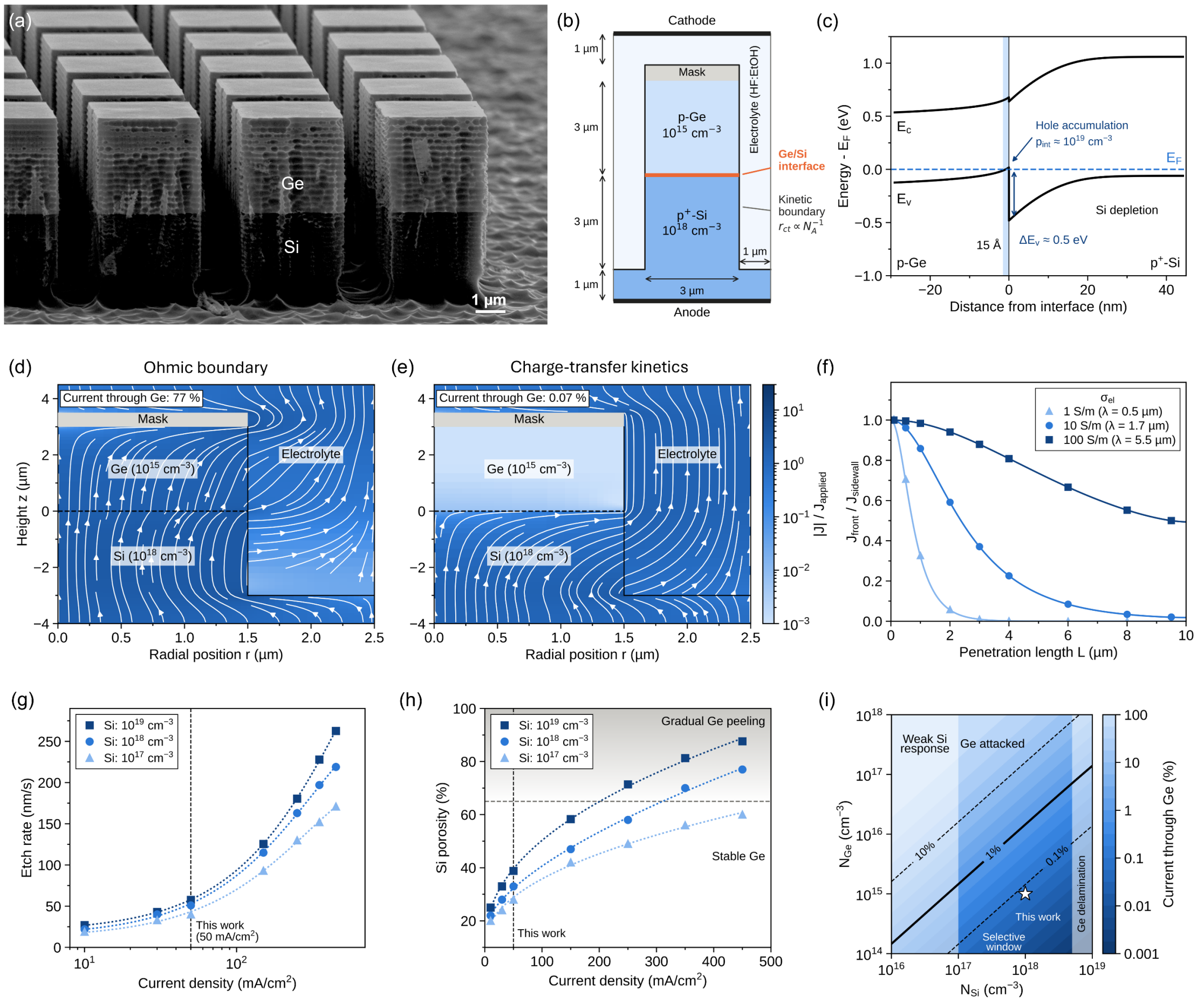

**Fig. 1.** Mechanism and process window for selective 2DFS formation. (a) SEM image of 3 × 3 μm² Ge/Si mesas. (b) Axisymmetric model geometry; charge-transfer kinetics with rct ∝ $N_A^{-1}$ apply at all semiconductor/electrolyte boundaries. (c) Equilibrium band diagram of the p-Ge/$p^+$-Si heterojunction, showing hole accumulation at the interface. (d,e) Simulated current-density distributions ($\sigma_{el}$ = 10 S $m^{-1}$) for (d) an ohmic boundary and (e) charge-transfer kinetics; the dashed line marks the Ge/Si interface. (f) Current density at the advancing 2DFS front, normalized to the sidewall value, versus penetration length for three electrolyte conductivities (20-μm mesa; lines: analytic solution, symbols: numerical model). (g) Si etch rate and (h) Si porosity versus current density for three Si doping levels; dotted lines are exponential-saturation (g) and allometric (h) fits ($R^2 \geq 0.997$), dashed vertical lines mark the operating current density, and the gradient in (h) marks gradual Ge peeling above ~65% porosity. (i) Simulated current through Ge versus Si and Ge doping; contours mark 0.1%, 1%, and 10%, shaded bands indicate the weak Si response and Ge delamination, and the star marks the operating point.

Cross-sectional STEM reveals the buried-interface transformation. The FIB-prepared lamella in Fig. 2(a) retains an intact crystalline Ge overlayer above porosified Si, consistent with the asymmetric electrochemical response expected from the doping design. High-resolution imaging [Figs. 2(b) and 2(c)] reveals a laterally continuous interfacial separation of approximately 15 Å. The gap is confined to the Ge/Si boundary and physically decouples the Ge layer from the substrate, rather than reflecting bulk consumption of Ge or uncontrolled porosification. EDX maps [Figs. 2(d)–2(g)] and line profiles [Figs. 2(h) and 2(i)] show an abrupt Ge/Si transition and oxygen enrichment near the interface. Atomic-resolution HAADF-STEM of the untreated reference shows a continuous lattice across the Ge/Si interface, with no resolved continuous amorphous interlayer (Sec. S5 of the supplementary material). The observed oxygen enrichment is consistent with oxidation of newly exposed interfacial surfaces after decoupling, although residual oxygen at the initial interface cannot be excluded. Importantly, the Ge region remains monocrystalline, indicating that the electrochemical transformation is localized at the buried defective interface while preserving the active epilayer. The approximately 15 Å separation is comparable to only a few atomic planes, yet it extends laterally over the imaged interface. This combination of nanometric thickness and lateral continuity is central to the 2DFS concept: it removes the structural connection to the misfit-dislocation-rich boundary without replacing the interface with a thick sacrificial cavity. The structural and compositional observations are therefore consistent with the proposed current-confinement mechanism presented in Fig. 1.

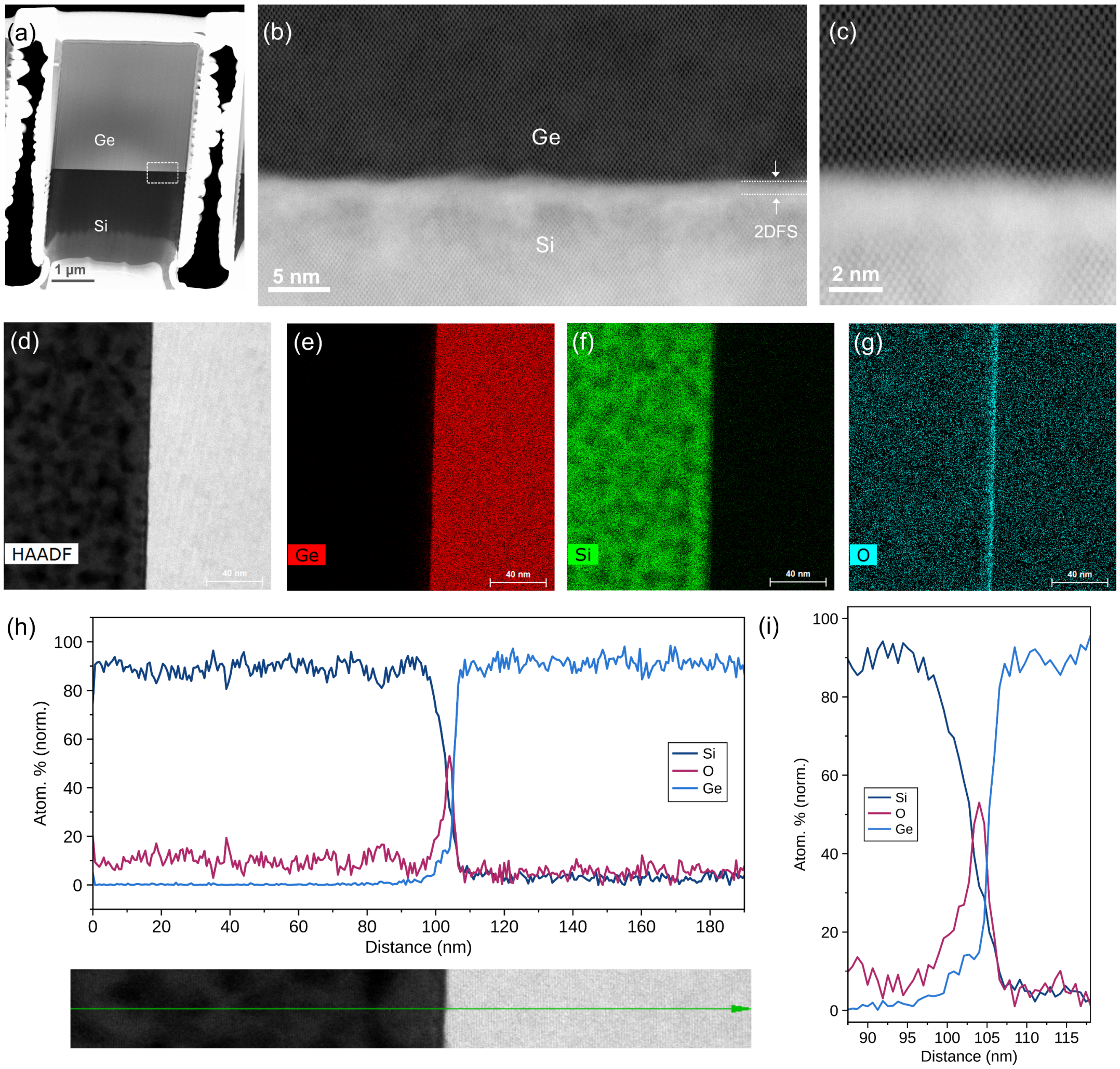

**Fig. 2.** Structure and chemistry of the buried 2DFS. (a) STEM image of a FIB lamella. (b,c) High- and atomic-resolution images showing a continuous approximately 15 Å Ge/Si separation. (d) HAADF-STEM image and EDX maps of (e) Ge, (f) Si, and (g) O. (h,i) Line profiles showing an abrupt Ge/Si transition and localized oxygen enrichment.

The resulting structure provides a free-surface-assisted pathway for defect evolution. Subsequent annealing drives dislocations toward nearby free surfaces and promotes recombination or annihilation through elastic-energy minimization.[26] The thin oxygen-containing layer may also stabilize the decoupled interface or hinder renewed dislocation propagation, as reported for oxide-mediated trapping in other heteroepitaxial systems;[36−38] this role, however, is not established by the present measurements. After etching, the samples were therefore softly annealed at 500 °C for 1 min in forming gas to promote structural reorganization without degrading the Ge layer. The patterned reference received the same soft anneal, after a prior cyclic anneal that did not significantly reduce its threading-dislocation density;[27] annealing alone therefore cannot account for the defect suppression *(Sec. S1 of the supplementary material)*. Before annealing, interfacial etching provides the geometric decoupling; during annealing, the reduction in substrate constraint supplies the mobility needed for the remaining defect segments to rearrange. The electrochemical and thermal steps therefore play complementary roles, with the first creating accessible free surfaces and the second enabling defect-energy minimization.

The effect of interfacial decoupling on local deformation was evaluated by geometric phase analysis (GPA) of high-resolution STEM images (Fig. 3 and *Sec. S6 of the supplementary material*).[39,40] In the reference Ge/Si sample [Figs. 3(a)–3(e)], the misfit-dislocation network is visible at the interface, and localized strain lobes are particularly evident in the $\varepsilon_{xx}$ map. Nonuniform deformation extends into the Ge region, consistent with a conventional relaxation pathway in which interfacial misfit dislocations generate TD-related distortions in the epilayer. By contrast, the 2DFS sample [Figs. 3(f)–3(j)] contains a structurally discontinuous interface and a more homogeneous deformation field across Ge. Localized signatures attributable to misfit-dislocation cores or TD propagation are not observed in the Ge layer; features within Si correspond instead to pores produced during electrochemical processing. The comparison indicates that the 2DFS does more than alter interface morphology: it reduces substrate constraint and changes how strain is accommodated after annealing, allowing the partly detached Ge crystal to reorganize toward nearby free surfaces. GPA across the porous, decoupled interface is interpreted qualitatively because bulk Si no longer provides a direct local lattice reference. Complementary 4D-STEM and virtual-ADF analysis of the same Ge/Si 2DFS system was reported previously.[27] The present GPA comparison is consistent with a Ge layer that is more relaxed and spatially more uniform after 2DFS formation than the patterned Ge/Si reference, which still accommodates mismatch through localized dislocation-mediated relaxation. The key difference is a transition from concentrated strain release at dislocation cores to a distributed, free-surface-assisted response that reduces the driving force for threading segments to reach the surface.

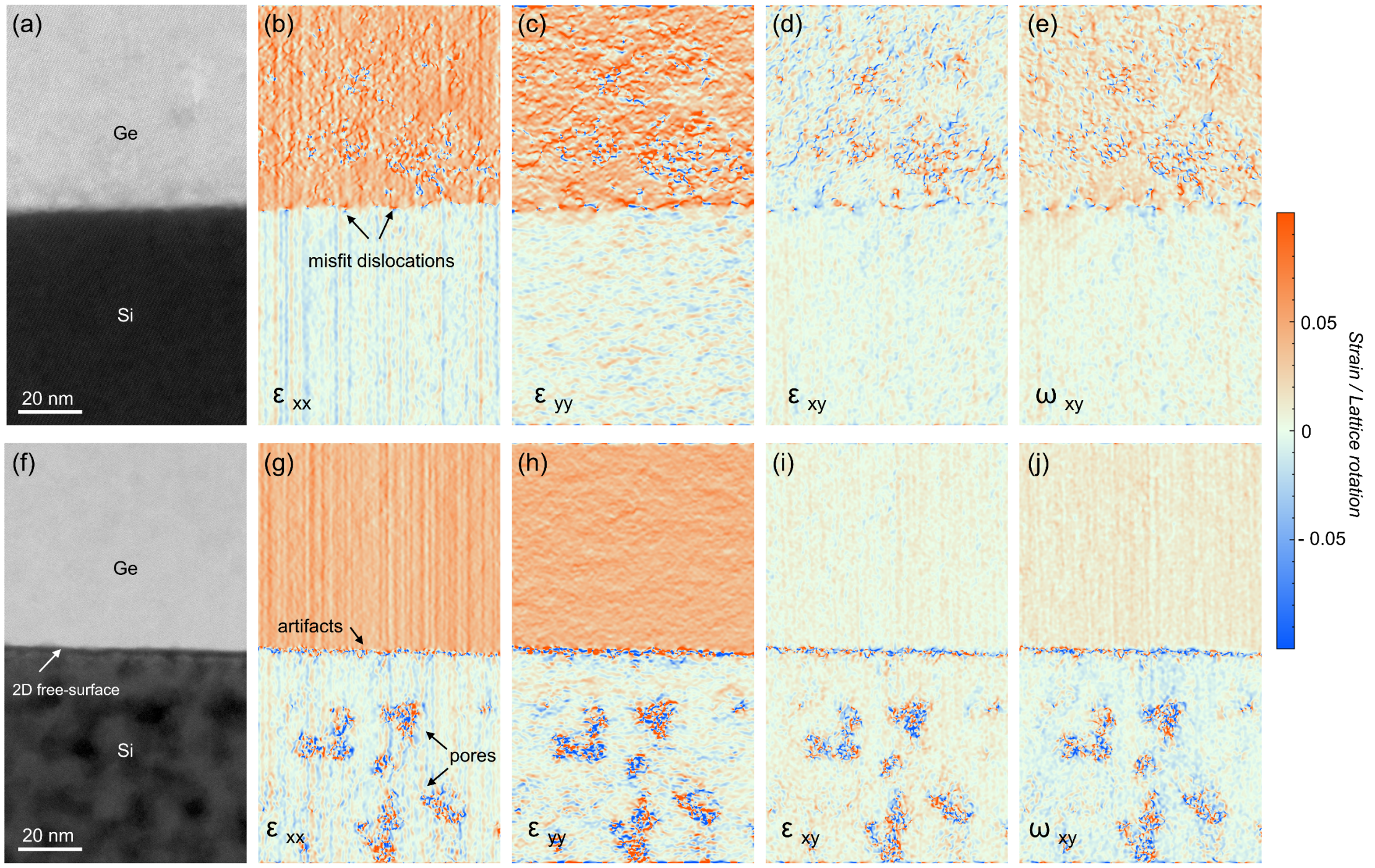

**Fig. 3.** GPA of reference (a–e) and 2DFS (f–j) Ge/Si interfaces. HRSTEM images (a,f) and corresponding $\varepsilon_{xx}$ (b,g), $\varepsilon_{yy}$ (c,h), $\varepsilon_{xy}$ (d,i), and $\omega_{xy}$ (e,j) maps show localized dislocation strain in the reference and a more uniform Ge deformation field after interfacial decoupling. Vertical lines in $\varepsilon_{xx}$ are scan-jump artifacts.

Plan-view TEM directly probes whether this modified relaxation pathway suppresses defect emergence near the Ge surface (Fig. 4). The reference sample contains multiple emergence sites [Fig. 4(a)], and high-resolution imaging near one site [Fig. 4(b)] shows contrast inhomogeneity associated with local lattice distortion. These observations agree with the nonuniform GPA maps and confirm that part of the interfacial dislocation content propagates through the Ge epilayer. In the 2DFS sample, no observable emergence sites are detected over the examined region [Fig. 4(c)], while the top Ge layers retain ordered lattice contrast [Fig. 4(d)]. Because plan-view TEM samples only a limited area, etch-pit analysis was performed on 36 mesas (four fields of nine 3 × 3 μm² mesas, ≈324 μm² in total). No etch pits were revealed on the 2DFS sample, corresponding to an upper bound of ≈$1 \times 10^6$ cm$^{-2}$ at the 95% confidence level, whereas the reference shows a TDD of ~$10^8$ cm$^{-2}$.[27] We therefore report this reduction in threading-dislocation density within the stated detection limits rather than claiming complete elimination of dislocations. Together, Figs. 3 and 4 show that the 2DFS decouples Ge from the original misfit-dislocation network and shifts relaxation from substrate-constrained deformation to free-surface-assisted reorganization and dislocation recombination during annealing. GPA reveals the resulting redistribution of strain at the buried interface, while plan-view TEM and etch-pit analysis show the corresponding suppression of defect emergence near the Ge surface. These complementary observations directly link interfacial decoupling to reduced threading-dislocation propagation through the Ge epilayer.

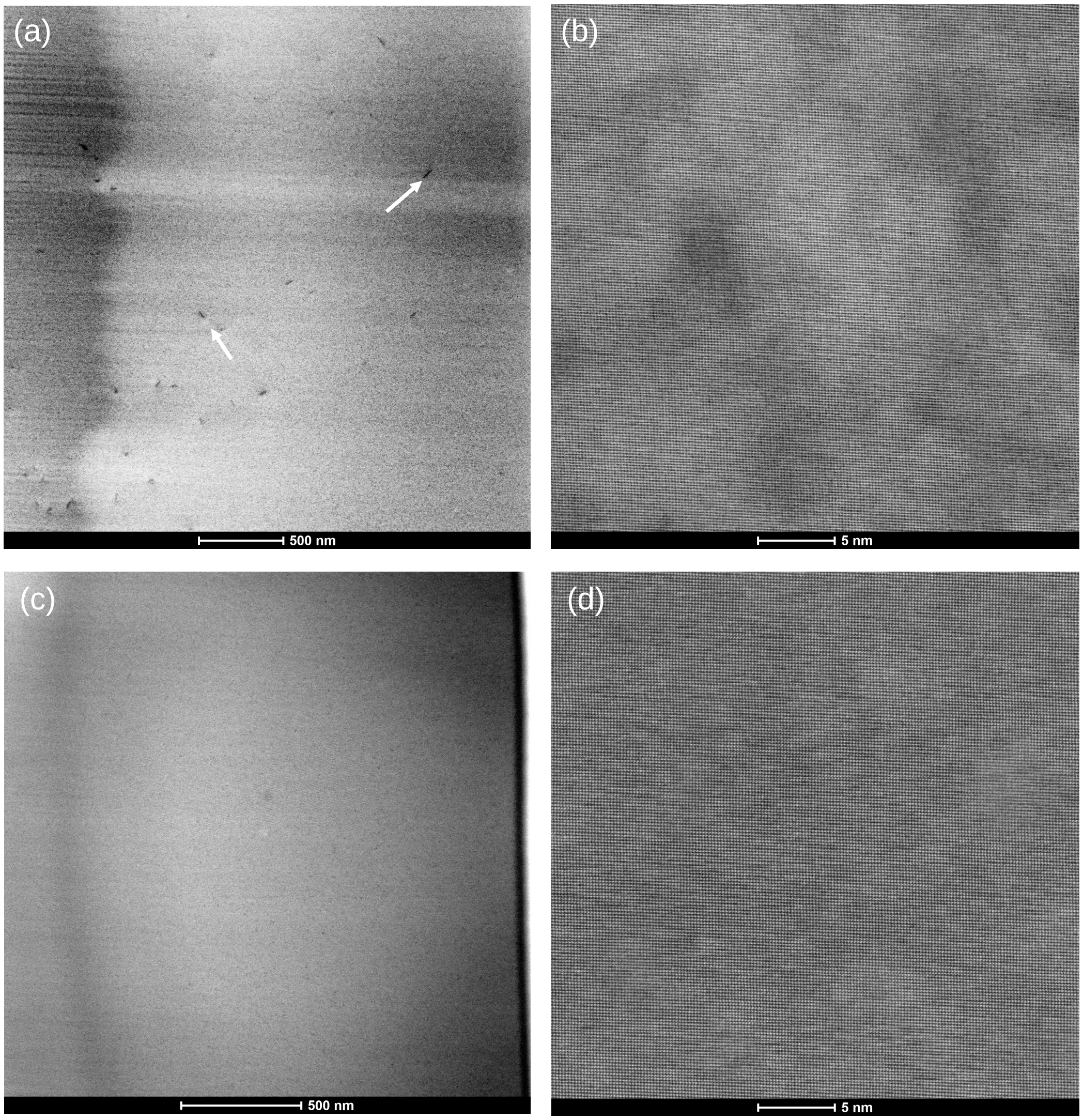

**Fig. 4.** Plan-view TEM of reference (a,b) and 2DFS (c,d) Ge/Si. The reference shows surface-emerging dislocations and local lattice distortion; the 2DFS sample shows no observable emergence sites in the examined region and preserves ordered top-Ge lattice contrast.

Raman spectroscopy quantifies the accompanying reduction in residual stress. Measurements used 532-nm excitation (measurement details in Sec. S7 of the supplementary material), for which the Ge probing depth is below approximately 20 nm,[41] and thus primarily sample the near-surface region. Values are given as mean ± standard deviation over more than 10 points per sample. Bulk Ge shows a Ge–Ge optical phonon peak at 300.84 ± 0.02 $cm^{-1}$ [Fig. 5(a)]. The patterned Ge/Si reference shifts to 300.40 ± 0.03 $cm^{-1}$, corresponding to approximately 0.10 ± 0.01% tensile strain using the conventional calibration,[42] whereas the 2DFS sample shifts toward bulk Ge to 300.65 ± 0.03 $cm^{-1}$, corresponding to approximately 0.05 ± 0.01%. The full width at half maximum increases from 1.30 ± 0.03 $cm^{-1}$ in bulk Ge to 1.92 ± 0.05 $cm^{-1}$ in the patterned reference, then narrows to 1.63 ± 0.04 $cm^{-1}$ after 2DFS formation, indicating reduced microstrain and structural disorder. Together with the Raman peak shift, this narrowing shows that the decoupled Ge layer retains only about half of the tensile strain measured in the patterned reference, while preserving the characteristic single Ge mode of high-quality (001) single-crystal material.[43] The small point-to-point dispersion indicates good reproducibility over the sampled positions, with a peak-position difference between the two samples exceeding the observed point-to-point variability.

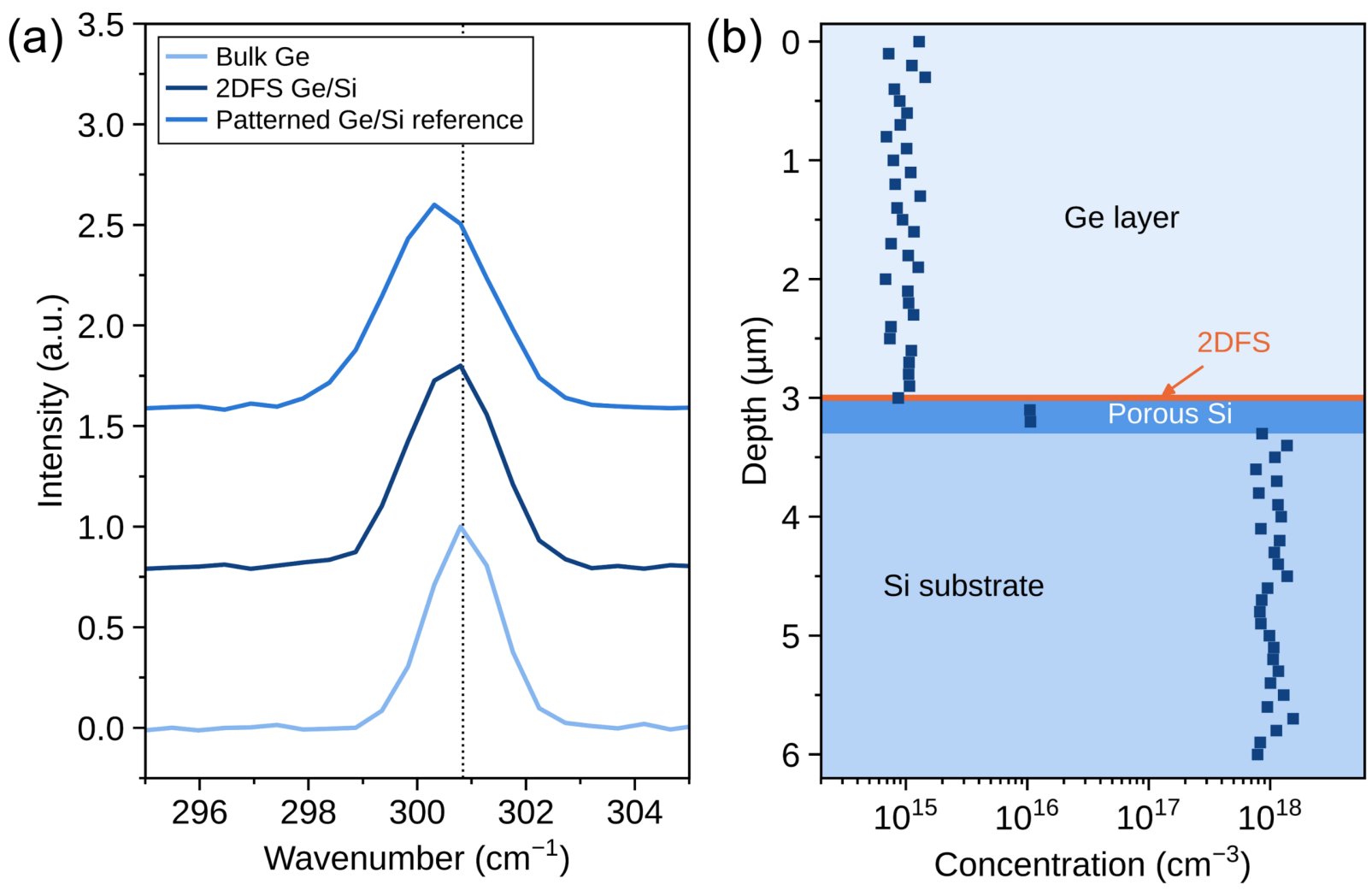

**Fig. 5.** (a) Raman spectra of bulk Ge, patterned Ge/Si, and 2DFS Ge/Si, showing reduced residual tensile strain after decoupling (vertically offset for clarity). (b) SRP carrier profile through Ge, the 2DFS/porous-Si region, and the Si substrate, confirming preservation of the Ge carrier concentration.

Spreading resistance profiling (SRP) was used to verify that selective interface transformation does not disrupt the vertical carrier profile [Fig. 5(b)]. Measurements were performed across a beveled region containing Ge, the 2DFS, porous Si, and crystalline Si. Within the Ge layer, the extracted carrier concentration remains close to the nominal $10^{15}$ $cm^{-3}$ value despite the porous region formed beneath it. The porous-Si region gives a lower apparent concentration of approximately $10^{16}$ $cm^{-3}$ because the nanostructured matrix increases resistance, reduces current-percolation pathways, and introduces geometrical contributions to the SRP response; this value is therefore not interpreted as the true dopant density. Deeper in the crystalline substrate, the profile recovers toward the nominal $10^{18}$ $cm^{-3}$ level. No strongly degraded region is detected within Ge. Together, Raman and SRP show that current-confined 2DFS formation reduces residual tensile strain while maintaining the intended electrical profile of the epilayer, complementing the structural evidence from Figs. 2–4. The analyzed bevel was approximately 20 × 20 µm², with an effective depth resolution of about 300 nm, sufficient to distinguish the main regions of the stack even though it cannot resolve the 2DFS thickness itself. The recovery of the nominal Si concentration below the porous layer further shows that the electrochemical modification remains localized rather than producing a deeply damaged substrate.

In summary, carrier-concentration engineering confines electrochemical current through hole-limited charge transfer at the Ge/Si heterointerface and enables controlled conversion of the misfit-dislocation-rich region into a laterally continuous nanometric free surface. The resulting 2DFS preserves the crystalline and electrical integrity of Ge, changes the strain-relaxation pathway, suppresses TD emergence after annealing, and reduces residual tensile strain. Unlike approaches that require modification during epitaxial growth, this post-epitaxial strategy directly transforms a buried defective interface and provides a general framework for defect and strain control in lattice-mismatched semiconductor heterostructures. The demonstrated link between current distribution, interfacial hole accumulation, electrochemical stability, atomic-scale decoupling, and post-anneal relaxation also provides practical design rules for extending buried free-surface engineering to other mismatched material systems.

**Acknowledgements**

Y. Bioud gratefully acknowledges support from the German Academic Exchange Service (DAAD).

**References.**

**SUPPLEMENTARY MATERIAL**

# Current-Confined Formation of a Buried Ge/Si Free Surface for Defect Filtering

Youcef A. Bioud [1*], Meriem Bouchilaoun [2], Waldemar Schreiber [1], Redouane Amrar [2,3], Gilles Patriarche [4], Ali Soltani [2,3], Stefan Janz [1*]

*[1] Division Photovoltaics, Fraunhofer Institute for Solar Energy Systems (ISE), Heidenhofstraße 2, 79110 Freiburg, Germany.*
*[2] Institut Interdisciplinaire d'Innovation Technologique (3IT), Université de Sherbrooke, 3000 Boulevard de l'université, Sherbrooke, J1K OA5 Québec, Canada*
*[3] Laboratoire Nanotechnologies Nanosystèmes (LN2) - IRL3463, CNRS, Université de Sherbrooke, INSA Lyon, École Centrale de Lyon, Université Grenoble Alpes, Sherbrooke, J1K 0A5 Québec, Canada.*
*[4] Université Paris-Saclay, CNRS, Centre de Nanosciences et de Nanotechnologies, 91120 Palaiseau, Paris, France.*

**Email: y.bioud@usherbrooke.ca; stefan.janz@ise.fraunhofer.de*

## S1. Ge/Si Mesa Array Fabrication and Selective Electrochemical Etching

Ge/Si mesa arrays were fabricated from commercially available Ge/Si wafers oriented along the ⟨100⟩ direction and comprising a 3-μm-thick Ge layer grown on Si. The heterostructures were p-type boron-doped, with nominal carrier concentrations of $10^{15}$ cm$^{-3}$ in Ge and $10^{18}$ cm$^{-3}$ in the Si substrate. The as-received material exhibited a threading-dislocation (TD) density on the order of $10^8$ cm$^{-2}$. Lateral patterning was carried out by electron-beam lithography to define square mesa arrays with lateral dimensions of 3 × 3, 5 × 5, 10 × 10, 15 × 15, and 20 × 20 μm². A 500-nm-thick positive-tone resist was used both as the lithographic mask and as a temporary protection layer for the subsequent wet electrochemical step. Electron-beam exposure was performed at 100 keV using a writing strategy adapted to improve feature fidelity, with lower resolution in the centre of the patterns and finer definition at the edges. After exposure, the resist was developed in methyl isobutyl ketone (MIBK) for 2 min. The mesas were then etched into the Ge/Si stack by inductively coupled plasma reactive ion etching using a deep reactive ion etching platform. The process relied on fluorine-based chemistry, with $SF_6$ as the main etching gas, $C_4F_8$ for sidewall passivation, and $O_2$ as an auxiliary gas to stabilize the process. Etching was performed in an alternating etch/passivation sequence following a Bosch-type approach in order to obtain high-aspect-ratio structures with controlled sidewalls. After completion of the main etch, an additional $SF_6$-only plasma step was applied to remove residual micromasking and improve the cleanliness of the etched surfaces. A final cleaning step was then used to remove plasma-induced deposits and preserve the integrity of the mesa structures before electrochemical processing. Selective electrochemical etching was performed in a two-electrode cell using a platinum wire as the counter electrode. The electrolyte consisted of a 5:1 mixture of hydrofluoric acid (HF, 49%) and anhydrous ethanol. The electrochemical treatment was carried out in bipolar pulsed mode[1,2] by applying alternating anodic and cathodic current densities of ±50 mA cm$^{-2}$, with a pulse duration of 1 s and a total treatment time of 1 min for the 3 × 3 μm² Ge/Si mesas. During electrochemical processing, the top surface of the mesas remained protected by the lithographic resist mask, thereby restricting the electrochemical transformation to the exposed regions and limiting direct attack of the Ge surface by the electrolyte. After etching, the resist mask was removed by immersion in acetone overnight. Post-etch thermal treatments were performed in forming gas ($N_2/H_2$, 90:10) using a rapid thermal annealing system (J.I.P. ELEC JetFirst) operated with a heating ramp of 10 °C s$^{-1}$. A soft anneal at 500 °C for 1 min was applied after selective electrochemical etching in order to assist the elimination of residual threading dislocations and promote structural reorganization of the decoupled Ge layer. The patterned Ge/Si reference was not left unannealed: it was first subjected to cyclic annealing at 800 °C following the conditions of Refs. 11 and 12, and then received the same soft anneal (500 °C, 1 min) as the 2DFS sample. As reported previously,[13] TEM performed before and after cyclic annealing showed no significant reduction of the TD density in the reference mesas, cyclic annealing being effective only in the vicinity of the mesa sidewalls.

## S2. Numerical Model of Current Distribution and Interfacial Reactivity

### S2.1 Model geometry, parameters, and boundary conditions

The current distribution during anodization was calculated with a steady-state, axisymmetric model of a single mesa unit cell [Fig. 1(b) of the main text], solved with a cell-centred finite-volume scheme on a non-uniform grid refined to 0.2 nm near the Ge/Si interface and the mesa edges. The unit cell comprises a 3-µm-diameter mesa made of a 3-µm-thick Ge layer on a 3-µm-high Si pedestal, a 1-µm electrolyte gap to the cell boundary, 1 µm of Si below the pedestal, a 0.5-µm-thick insulating resist mask on top of Ge, and 1 µm of electrolyte above the mask. The back of the Si (anode) and the top of the electrolyte (cathode) were held at fixed potentials, and all other outer boundaries are symmetry (zero-flux) planes. Because all constitutive relations are linear, the relative current distribution is independent of the applied potential and therefore also describes the galvanostatic pulses used experimentally. Results are reported as the fraction of the total anodic current that leaves through Ge surfaces, $\eta_{Ge}$.

Bulk conductivities were calculated from the nominal boron concentrations, $\sigma = qN_A\mu_p$, using the Masetti mobility model for Si[8] and a hole mobility of ≈1840 $cm^2\ V^{-1}\ s^{-1}$ for low-doped Ge. This gives $\sigma_{Ge} \approx 30$ S $m^{-1}$ for $10^{15}$ $cm^{-3}$ and $\sigma_{Si} \approx 2.4 \times 10^3$ S $m^{-1}$ for $10^{18}$ $cm^{-3}$, with $4.7 \times 10^2$ and $1.1 \times 10^4$ S $m^{-1}$ for $10^{17}$ and $10^{19}$ $cm^{-3}$ substrates. The electrolyte conductivity was set to $\sigma_{el} = 10$ S $m^{-1}$ and varied between 1 and $10^4$ S $m^{-1}$. At every semiconductor/electrolyte boundary, charge transfer was described by a linearized kinetic law, $j = \Delta\varphi/r_{ct}$, in series with the bulk resistances. Because anodic dissolution of p-type Ge and Si consumes holes supplied at the surface,[2,6] the charge-transfer resistance was scaled inversely with the acceptor concentration:

$$r_{ct} = r_{ct,ref}\ (10^{18}\ cm^{-3} / N_A),$$

with $r_{ct,ref} = 2\ \Omega\ cm^2$ and the same prefactor for Ge and Si. For comparison, the purely ohmic description of earlier current-distribution simulations of Si electrochemical etching[3,4] was also evaluated, i.e., continuity of potential at the semiconductor/electrolyte boundary with $\sigma_{el} = 10^4$ S $m^{-1}$, $\sigma_{Ge} = 2$ S $m^{-1}$, a Si substrate of 10 S $m^{-1}$ [Si(b)], and a 1-nm interfacial Si layer of $10^2$ S $m^{-1}$ [Si(a)].

## S2.2 Origin of current confinement and sensitivity analysis

With an ohmic boundary and realistic conductivities, 77% of the anodic current would leave through the Ge sidewall [Fig. 1(d) of the main text], because $\sigma_{Ge}$ exceeds $\sigma_{el}$. In an ohmic model, Ge becomes passive ($\eta_{Ge} < 1\%$) only for $\sigma_{el}$ $^{-}$ $3 \times 10^3$ S m $^{1}$ (Fig. S1b). With charge-transfer kinetics, $\eta_{Ge}$ falls to 0.07% [Fig. 1(e) of the main text] and becomes independent of $\sigma_{el}$ between 1 and $10^4$ S $m^{-1}$ (Fig. S1b).

Varying the conductivity of the 1-nm Si(a) layer from 1 to 100 times that of the substrate leaves $\eta_{Ge}$ unchanged in the kinetic model (0.07%) and changes it only from 0.058% to 0.052% in the original ohmic model (Fig. S1a). The current carried laterally by this layer is below 0.002% of the total in the original model and below 0.5% in the kinetic model even for a 100-fold conductivity. The predicted confinement therefore does not rely on the assumed layer, which was removed from the reference model. $\eta_{Ge}$ is likewise insensitive to the bulk conductivities: changing $\sigma_{Ge}$ by a factor of 0.25–4 and $\sigma_{Si}$ by 0.5–2, or the substrate thickness from 1 to 20 µm, leaves $\eta_{Ge}$ at 0.07%, because the charge-transfer resistance dominates over the bulk resistance at the micrometre scale. The relevant inputs are thus the surface hole concentrations, which spreading resistance profiling confirms to be ~$10^{15}$ $cm^{-3}$ in Ge and ~$10^{18}$ $cm^{-3}$ in Si [Fig. 5(b) of the main text].

**(a)** **(b)**

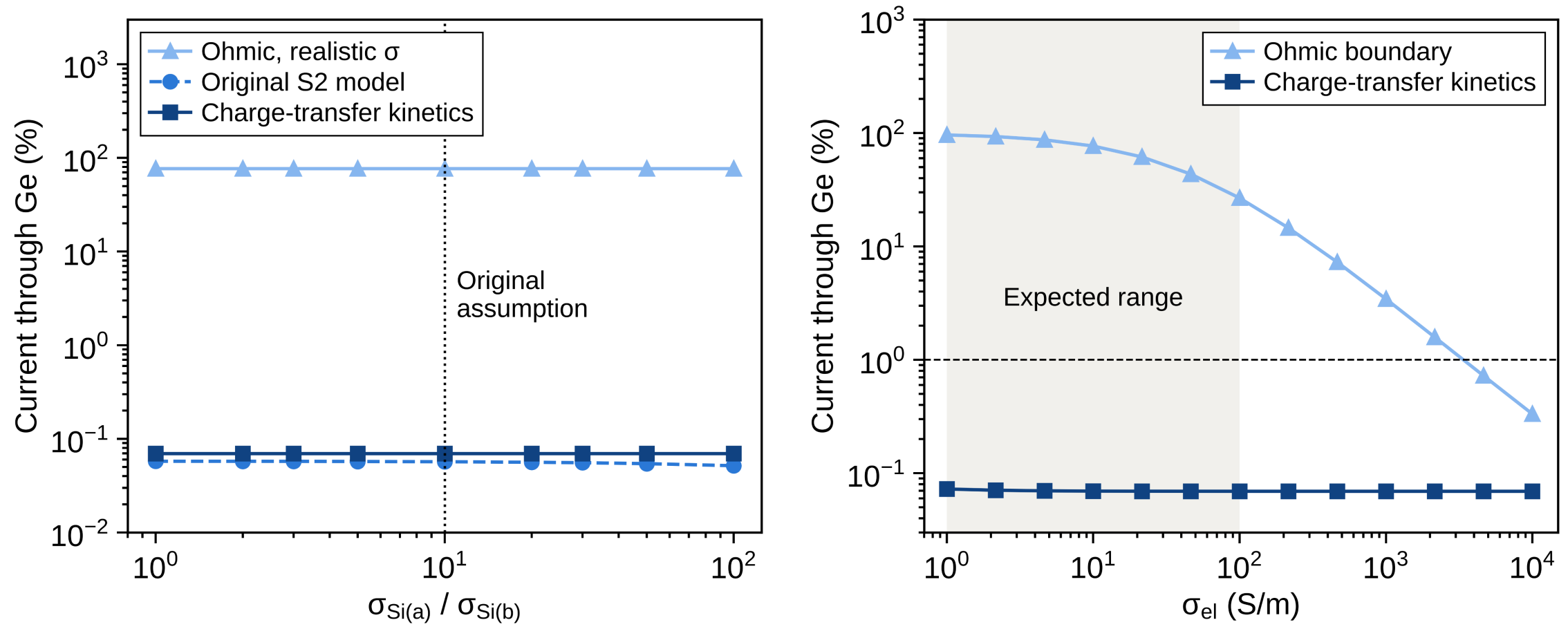

**Fig. S1.** (a) Fraction of the anodic current leaving through Ge, $\eta_{Ge}$, versus the conductivity ratio of a 1-nm interfacial Si(a) layer to the Si substrate, for the original ohmic model, an ohmic model with realistic conductivities ($\sigma_{el}$ = 10 S $m^{-1}$), and the charge-transfer model. (b) $\eta_{Ge}$ versus electrolyte conductivity for ohmic and charge-transfer boundaries; the shaded band marks the expected range for HF/ethanol electrolytes, and the dashed line marks 1%.

Fig. S2(a) shows $\eta_{Ge}$ versus Si doping for three Ge doping levels. Keeping the fraction of anodic current leaving through Ge below 1% requires $N_{Si}/N_{Ge} \gtrsim 60$. For $10^{15}$ $cm^{-3}$ Ge, the operating substrate ($10^{18}$ $cm^{-3}$) gives $\eta_{Ge}$ = 0.07%, whereas $10^{17}$ and $10^{19}$ $cm^{-3}$ give 0.7% and 0.007%, respectively; the full map is shown in Fig. 1(i) of the main text. Fig. S2(b) tests the kinetic assumptions. $\eta_{Ge}$ remains ≈0.07% for $r_{ct,ref} \geq 0.1\ \Omega\ cm^2$, increases only when $r_{ct,ref}$ approaches the bulk-resistance scale, and stays below 1% down to $10^{-3}\ \Omega\ cm^2$. Planar and axisymmetric geometries give nearly identical results. A weaker doping dependence, $r_{ct} \propto N_A^{-1/2}$, raises $\eta_{Ge}$ to ≈2%, which remains far below the ohmic limit.

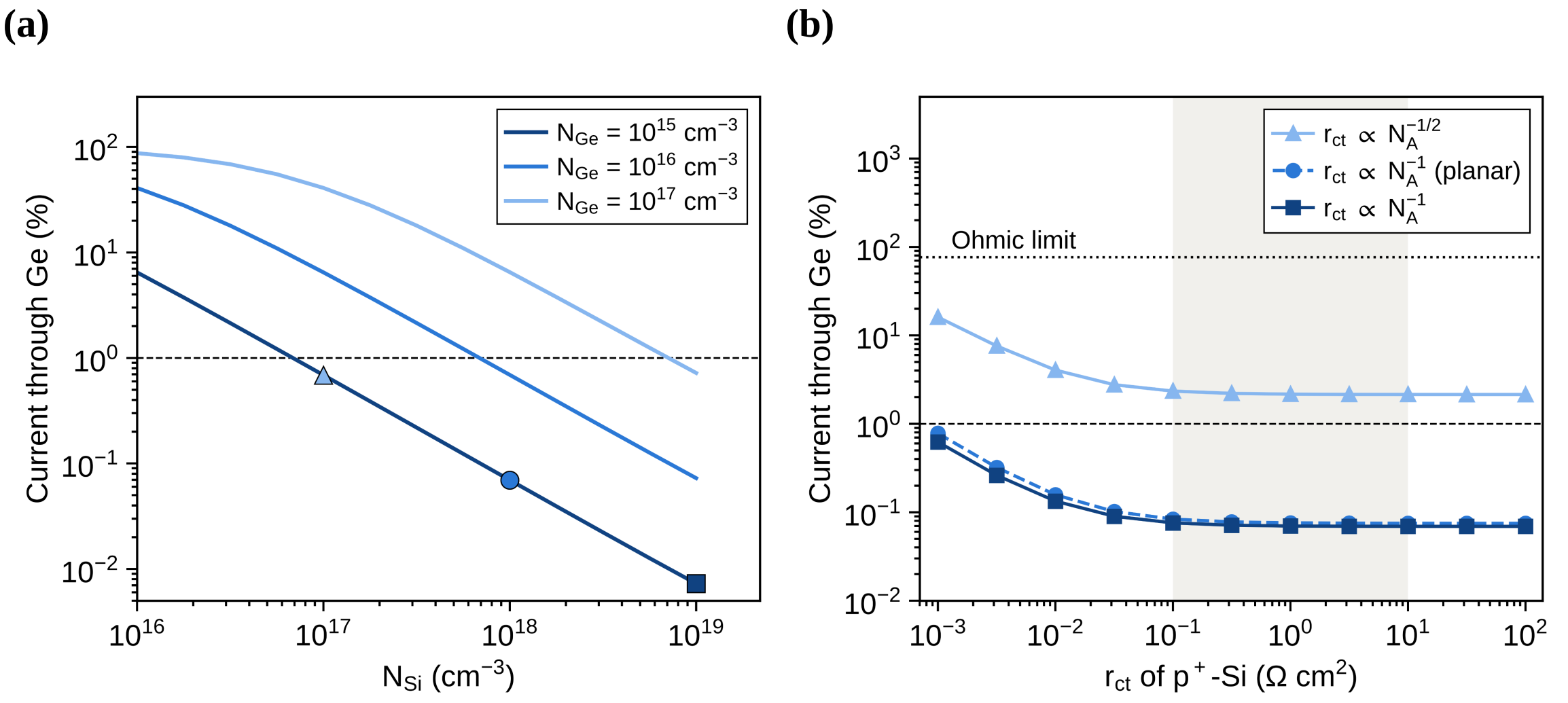

**Fig. S2.** (a) $\eta_{Ge}$ versus Si substrate doping for three Ge doping levels; symbols mark the Si doping levels used experimentally ($10^{17}$, $10^{18}$, and $10^{19}$ $cm^{-3}$), and the dashed line marks 1%. (b) $\eta_{Ge}$ versus the charge-transfer resistance of p$^+$-Si for $r_{ct} \propto N_A^{-1}$ (axisymmetric and planar geometries) and $r_{ct} \propto N_A^{-1/2}$; the dotted line is the ohmic limit, and the shaded band is the assumed range of $r_{ct,ref}$.

## S2.3 Lateral propagation of the 2DFS front and mesa-size design rule

To describe propagation after the interface opens, the model includes an electrolyte-filled interfacial gap of thickness h = 1.5 nm and length L entering from the mesa sidewall along the Ge/Si interface. The current feeding the front must flow through this thin gap, whose lateral resistance competes with charge transfer at the gap walls. For a radial gap in a mesa of radius R, the current density at the front normalized to that at the sidewall is

$$J_{front}/J_{side} = (1/a) / [I_0(b)K_1(a) + K_0(b)I_1(a)], \quad a = (R - L)/\lambda, \quad b = R/\lambda, \quad \lambda = (\sigma_{el} h r_{ct})^{1/2},$$

where $I_n$ and $K_n$ are modified Bessel functions. The numerical model reproduces this solution [Fig. 1(f) of the main text]. For $\sigma_{el} = 10\ S\ m^{-1}$ and $r_{ct} = 2\ \Omega\ cm^2$, λ ≈ 1.7 μm: the centre of a 3-μm mesa still receives ≈84% of the sidewall current density, whereas this value drops to ≈20% for a 10-μm mesa and ≈2% for a 20-μm mesa. Requiring at least 50% of the sidewall current density at the mesa centre gives a maximum mesa width $w_{max} \approx 3.6\lambda$ (Fig. S3), i.e., ≈3–14 μm for $r_{ct} = 0.5$–$10\ \Omega\ cm^2$ at $\sigma_{el} = 10\ S\ m^{-1}$. Transport through the electrolyte-filled porous Si beneath the 2DFS increases the effective gap thickness (dashed line, $h_{eff}$ = 15 nm), so these values are lower bounds. An interfacial reactivity enhancement k (Sec. S2.4) multiplies the front current density by the same factor.

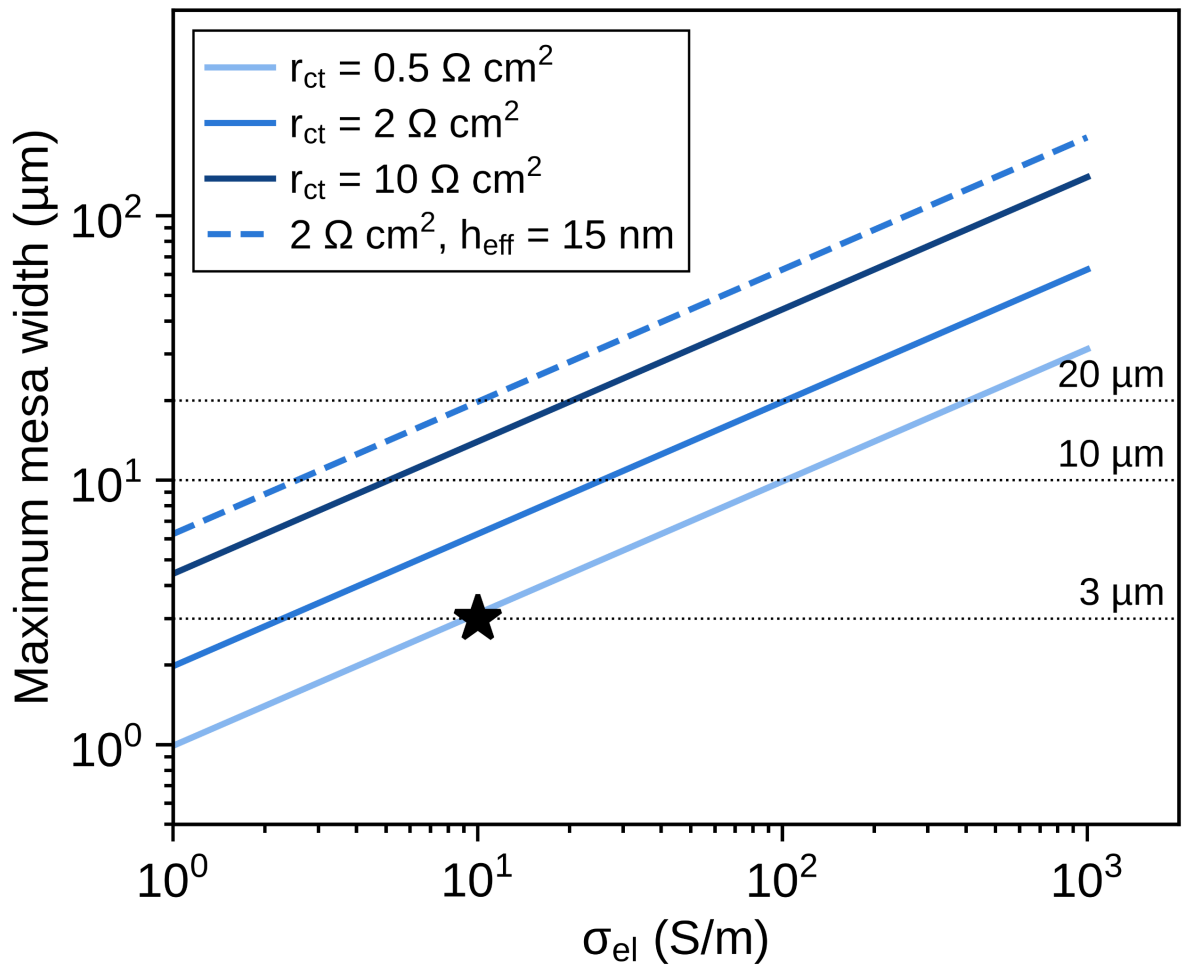

**Fig. S3.** Maximum mesa width for complete decoupling, defined by a front current density at the mesa centre of at least 50% of the sidewall value, versus electrolyte conductivity for different charge-transfer resistances and effective gap thicknesses. Dotted lines mark 3-, 10-, and 20-µm mesas; the star marks the 3-µm mesas used here with $\sigma_{el}$ = 10 S m$^{-1}$.

### S2.4 Hole accumulation at the p-Ge/p$^+$-Si heterojunction

The equilibrium band diagram of the heterojunction was obtained by solving the one-dimensional Poisson equation at 300 K with Fermi–Dirac statistics (Aymerich-Humet approximation of the Fermi integral $F_{1/2}$),[9] using the effective densities of states, band gaps, and permittivities of Ge and Si, fully ionized acceptors, and a valence-band offset $\Delta E_v \approx 0.5$ eV estimated from the electron-affinity rule. $\Delta E_v$ was varied between 0.3 and 0.7 eV. An equilibrium treatment is justified because only ~0.07% of the anodic current crosses the junction toward Ge.

The offset transfers holes from p$^+$-Si into Ge and produces an accumulation layer with $p_{int} \approx 1.0 \times 10^{19}$ cm$^{-3}$ at the interface, about ten times the substrate concentration, and a ~20-nm depletion region on the Si side [Fig. 1(c) of the main text and Fig. S4(a)]. The interfacial hole enrichment $p_{int}/N_{Si}$, taken as a proxy for the local reactivity enhancement k ($r_{ct} \propto 1/p$), lies between 6 and 15 for $N_{Si} = 10^{18}$ cm$^{-3}$ (Fig. S4b). This is comparable to the k ≈ 10 required for the front to advance about eight times faster than the surrounding Si is porosified (Sec. S2.3). The Ge layer whose hole density exceeds that of the substrate is ≈5 nm thick for $10^{18}$ cm$^{-3}$ and decreases to ≈1.7 nm for $10^{19}$ cm$^{-3}$ (Fig. S4c), whereas its dependence on the Ge doping is weak (4.9–5.5 nm for $10^{15}$–$10^{17}$ cm$^{-3}$). This self-limiting reactive layer is consistent in order of magnitude with the ~15 Å separation measured by STEM; quantum confinement of holes and misfit-dislocation states, which are not included, would further reduce its effective thickness. The calculation also predicts that the 2DFS thickness decreases with increasing substrate doping.

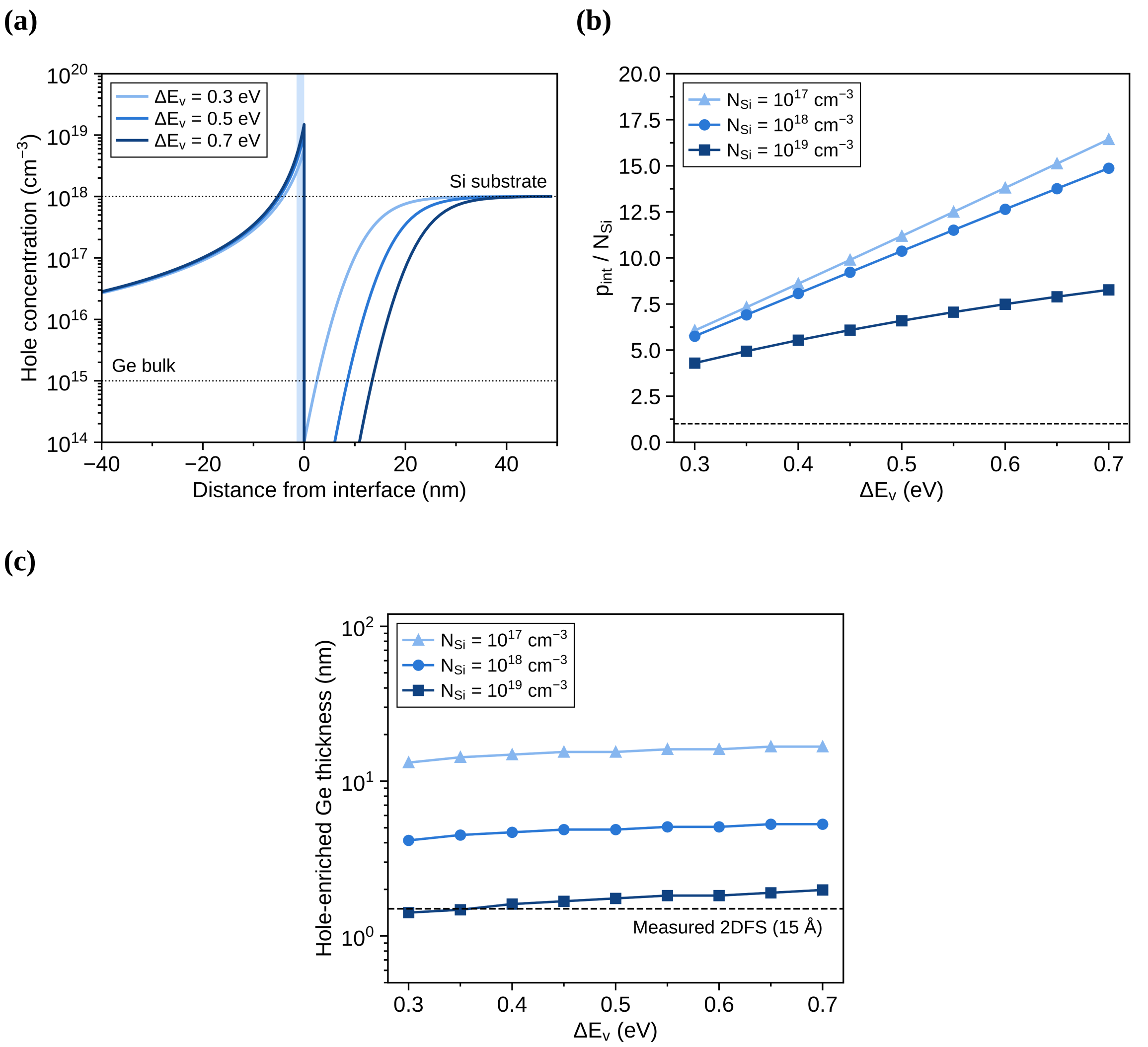

**Fig. S4.** (a) Hole concentration across the p-Ge ($10^{15}$ cm$^{-3}$)/p$^+$-Si ($10^{18}$ cm$^{-3}$) heterojunction for $\Delta E_v$ = 0.3–0.7 eV; the shaded band marks the 15 Å 2DFS. (b) Interfacial hole enrichment $p_{int}/N_{Si}$ and (c) thickness of the Ge layer with $p \geq N_{Si}$ versus $\Delta E_v$ for three Si doping levels; the dashed line in (b) marks $p_{int} = N_{Si}$, and that in (c) marks the measured 2DFS thickness.

### S2.5 Model limitations

The model describes the initial, steady-state current distribution with linear constitutive relations. It does not describe pore nucleation, porous-Si growth, or the time evolution of the front, and it uses a linearized charge-transfer law with the same kinetic prefactor for Ge and Si. Because the electrolyte conductivity and $r_{ct,ref}$ were not measured, they were varied over wide ranges: the conclusions of Sec. S2.2 hold over the whole range, whereas the absolute mesa widths of Sec. S2.3 depend on their values. The Poisson calculation neglects interface and dislocation states, strain, and Si–Ge intermixing, and the electron-affinity estimate of $\Delta E_v$ is approximate.

### S3. Porosity Determination by Infrared Reflectance Measurements

The porosity of the porous Si layers was estimated from infrared reflectance measurements. When infrared light is incident on freshly prepared porous Si, partial reflection occurs both at the air/porous-Si interface and at the porous-Si/crystalline-Si interface. Interference between these reflected beams produces Fabry–Pérot fringes in the reflectance spectrum. The fringe pattern was used to extract the effective refractive index of the porous layer according to:

$$n_{PSi} = m / (2\Delta \nu t)$$

where $m$ is the number of interference fringes within the spectral range $\Delta\nu$, and $t$ is the porous-layer thickness, determined independently from SEM cross-sections.

Assuming an isotropic effective medium, the porosity was then calculated using the Landau–Lifshitz–Looyenga model [5]:

$$P = (n_{PSi}^{2/3} - n_{Si}^{2/3}) / (1 - n_{Si}^{2/3})$$

where the refractive index of air was taken as unity and the refractive index of crystalline Si was set to $n_{Si} \approx 3.42$ in the spectral range considered. This approach provides an effective estimate of porous-layer volume fraction by relating the measured optical response to the relative contributions of air and Si in the nanostructured medium.

### S4. Electrochemical Dissolution Pathways and Interfacial Selectivity

The electrochemical reactions involved are consistent with known Ge and Si dissolution pathways in HF-based media. In Ge, anodic dissolution can proceed through divalent or tetravalent pathways depending on the applied potential [2]:

$$Ge + 6HF + h^+ \rightarrow GeF_6^{2-} + 4H^+ + H_2 + e^-$$

$$Ge + 6HF + 4h^+ \rightarrow GeF_6^{2-} + 6H^+$$

For Si, dissolution can be described through the sequence [6]:

$$Si + 2HF + nh^+ \rightarrow SiF_2 + 2H^+ + (2 - n)\, e^-$$

$$SiF_2 + 2HF \rightarrow SiF_4 + H_2$$

$$SiF_4 + 2HF \rightarrow H_2SiF_6$$

where $h^+$ is a hole, $e^-$ an electron, and n the number of exchanged charges. Under the selected conditions, the low-doped Ge behaves as a relatively passive overlayer, whereas the more conductive p-type Si sustains most of the electrochemical current. Combined with the enhanced reactivity of the defective interface, this asymmetry concentrates dissolution beneath the Ge layer and along the dislocation-rich heterointerface.

**S5. Structure of the Ge/Si Interface in the Annealed Reference**

The sample examined in Fig. S5 is the patterned Ge/Si reference without electrochemical etching, rather than an as-grown sample. It underwent cyclic annealing followed by the same soft anneal applied to the 2DFS sample (500 °C, 1 min, forming gas). The reference was examined by HAADF-STEM on a FIB lamella (Titan, 200 kV, ⟨110⟩ zone axis). As shown in Fig. S5, the Ge and Si atomic columns are continuous across the interface, and no continuous amorphous interlayer is resolved in the imaged region. The HAADF intensity averaged along the interface decreases monotonically from the Ge level to the Si level over ≈3.5 nm; this apparent transition width can be influenced by interface undulation within the field of view and projection through the lamella thickness. No intensity minimum below the Si level is detected. These observations support an epitaxial interface but do not exclude trace or sub-monolayer oxygen at the interface of the annealed reference. The oxygen enrichment detected in the 2DFS sample (Fig. 2 of the main text) is consistent with oxidation following electrochemical decoupling, although its origin and timing cannot be established conclusively from this structural comparison.

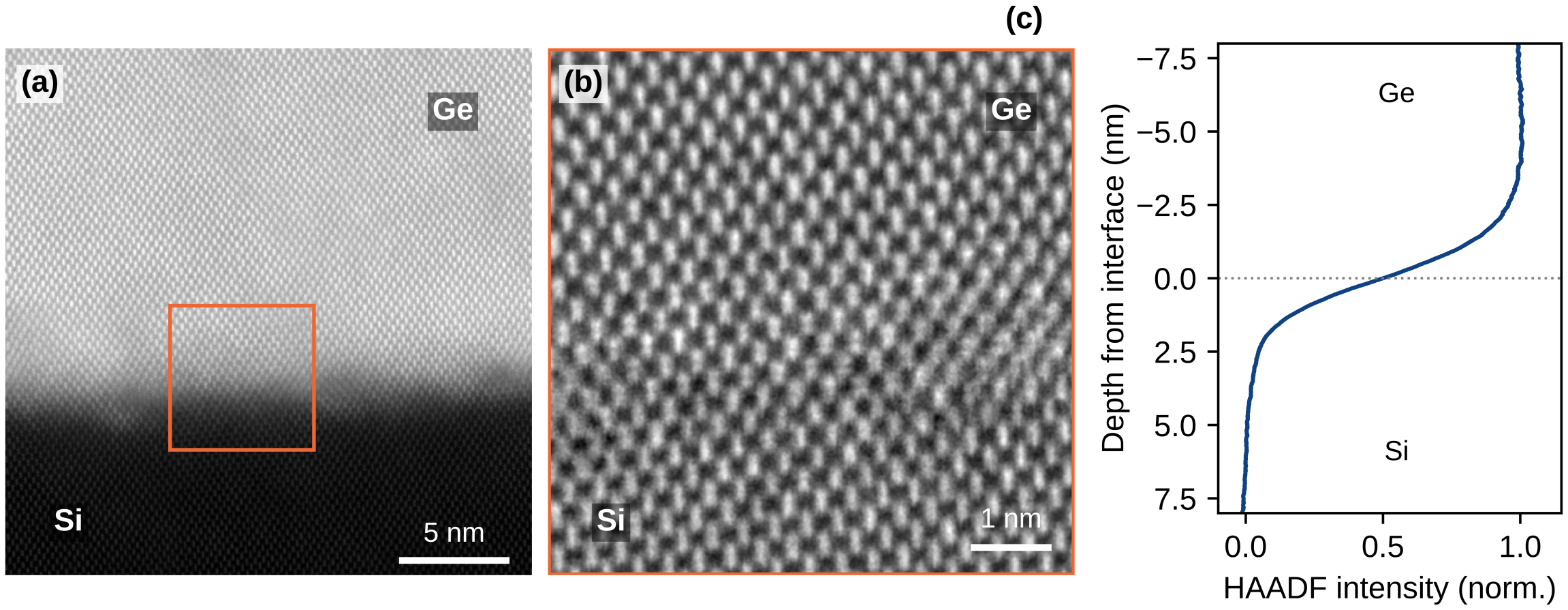

**Fig. S5.** Ge/Si interface of the annealed patterned reference (without electrochemical treatment). (a) Atomic-resolution HAADF-STEM image showing Ge (top) and Si (bottom); the box marks the region magnified in (b). (b) Magnified interface region after background normalization (subtraction of a Gaussian-smoothed background, σ ≈ 0.3 nm, division by the local contrast, and Gaussian smoothing, σ ≈ 0.014 nm), which displays the Ge and Si atomic columns with comparable contrast; the lattice is continuous across the interface. (c) HAADF intensity averaged along the interface over the full field of view in (a), normalized between the Si (0) and Ge (1) levels, as a function of depth from the interface (50% intensity point).

## S6. Geometric Phase Analysis of Strain at the Ge/Si Interface

Strain mapping at the Ge/Si interface was performed by geometric phase analysis (GPA) using high-resolution STEM images, following the methodology described in ref. [7]. Two intense {111} diffraction spots were selected in reciprocal space to reconstruct the local displacement field, from which the in-plane strain components $\varepsilon_{xx}$, $\varepsilon_{yy}$, and $\varepsilon_{xy}$, as well as the lattice rotation $\omega_{xy}$, were obtained. The calculations were carried out in the image coordinate system, with the reference reciprocal lattice vectors defined from the full field of view. GPA was used here to compare the spatial distribution of strain at the Ge/Si interface in the reference and 2DFS samples and to identify local strain heterogeneities associated with defect structures.

## S7. Raman Spectroscopy Measurement Parameters

Raman spectra were acquired at room temperature in backscattering geometry with 532-nm excitation using a HORIBA LabRAM HR Evolution micro-Raman spectrometer equipped with a 100× (NA 0.9) objective (spot diameter ≈ 0.8 µm). The laser power at the sample was 0.5 mW, the acquisition time was 30 s with 3 accumulations, and a 1800 lines $mm^{-1}$ grating was used. The spectrometer was calibrated against the Si phonon peak at 520.7 $cm^{-1}$, and the laser power was kept low and identical for all samples to avoid laser-induced heating. For each sample, more than 10 spectra were acquired at different positions. Each spectrum was fitted with a Lorentzian function, and the reported peak positions and full widths at half maximum are the mean ± standard deviation over all measured points. Strain values were obtained from the peak shift using the calibration of Ref. 10, and their uncertainties by dividing the standard deviation of the peak position by the strain-shift coefficient.